\documentclass[twocolumn,aps,prb,showpacs,floats,amsmath,amssymb]{revtex4}
\usepackage[dvips]{graphicx}
\usepackage{dcolumn}
\usepackage{bm}

\begin{document}


\title{Tunability of the dielectric response of epitaxially strained
SrTiO$_3$ from first principles}

\author{Armin Antons}
\email{a.antons@fz-juelich.de}
\altaffiliation{Present address: Institut of Solid State Research (IFF), 
Research Centre J{\"u}lich, 52425 J{\"u}lich, Germany.}
\author{J. B. Neaton}
\altaffiliation{Present address: The Molecular Foundry, Materials Sciences
                Division, Lawrence Berkeley National Laboratory, 
                Berkeley, CA 94720, USA.}
\author{Karin M. Rabe}
\author{David Vanderbilt}
\affiliation{Department of Physics and Astronomy, Rutgers University,\\
Piscataway, New Jersey 08854-8019}
\date{July 2, 2004}

\begin{abstract}
The effect of in-plane strain on the nonlinear dielectric
properties of SrTiO$_3$ epitaxial thin films is calculated using
density-functional theory within the local-density approximation.
Motivated by recent experiments, the structure, zone-center
phonons, and dielectric properties with and without an external
electric field are evaluated for several misfit strains within
$\pm$3\% of the calculated cubic lattice parameter. In these
calculations, the in-plane lattice parameters are fixed, and all
remaining structural parameters are permitted to relax. The
presence of an external bias is treated approximately by applying a
force to each ion proportional to the electric field.  After
obtaining zero-field ground state structures for various strains, 
the zone-center phonon
frequencies and Born effective charges are computed, yielding
the zero-field dielectric response.
The dielectric response at finite electric field bias is obtained
by computing the field dependence of the structure and polarization
using an approximate technique.
The results are compared with recent experiments and a previous
phenomenological theory. The tunability is found to be strongly
dependent on the in-plane lattice parameter, showing markedly
different behavior for tensile and compressive strains. Our results
are expected to be of use for isolating the role of strain in
the tunability of real ultrathin epitaxial films.
\end{abstract}


\maketitle

\section{\label{sec:level1}INTRODUCTION\protect\\}

Microwave dielectric materials strongly
tunable by the application of an electric bias field are
increasingly important for a variety of applications in microwave
electronics, including tunable capacitors and oscillators, phase
shifters, and delay lines.\cite{Fiedziuszko1,Varadan1,Horwitz1} The
dependence of dielectric constant $\epsilon$ on bias field $\mathcal{E}$
is typically strongly non-linear, and a tunability parameter $n$
can be defined as
$n=[\epsilon(0)-\epsilon(\mathcal{E})]/\epsilon(0)$, where
$\mathcal{E}$ is an operational bias field of interest.  Because of
the potential technological impact, considerable
experimental effort has been directed toward the design, control,
and optimization of highly tunable low-loss materials.

For many applications, thin-film morphologies are required, and
in recent years the growth of high-quality ultrathin
perovskite films with unprecedented atomic-level control has become possible, using
techniques such as molecular-beam epitaxy (MBE) \cite{Yoneda1}
and pulsed-laser deposition (PLD). \cite{Qing1}
These efforts have achieved the growth of
single-crystal films of nanometer-scale thickness with a minimum of
defects.
Even so, the dielectric properties of these films often differ quite
substantially from bulk\cite{gim,knauss}. This difference is due to a number of
factors, of which the strain in the film is in many cases among
the most important.
More specifically, the mechanical
boundary conditions on a coherent epitaxial film require the 
in-plane lattice constant of the
film to stretch (or contract) to match the lattice constant of the
substrate. Both experimental and
theoretical studies have found that even small epitaxial strains
can appreciably influence
the Curie temperature and the dielectric permittivity and tunability of
BaTiO$_3$, SrTiO$_3$, and Ba$_{1-x}$Sr$_x$TiO$_3$ (BST) thin films.
\cite{ramesh1,Pertsev1,James1,Park1}

Paraelectric BST at $x$ near 0.5 is already known to possess
a large dielectric response and high
tunability at room temperature, associated with close proximity to the
ferroelectric phase transition at about $-23$ K (Ref.~\onlinecite{Ban1}).
However, significantly lower loss has been reported
for single-crystal SrTiO$_{3}$ (STO) thin films,\cite{Li1} possibly
due to the absence of compositional disorder.
The ground state of STO is nonpolar but nearly
ferroelectric, and thus small applied stresses would be expected to
have a significant influence on the Curie temperature and associated
susceptibilities. Indeed, recent experimental studies\cite{Hyun1,Haeni1} 
indicate that significant changes in Curie temperature and dielectric 
properties occure when STO is grown epitaxially on substrates with
different lattice constants.

One of the best-studied perovskite materials, pure STO is the subject of
considerable experimental and theoretical literature.
STO adopts the centrosymmetric cubic perovskite structure at room temperature,
and undergoes a structural phase transition from the cubic to a
tetragonal, nonpolar antiferrodistortive (AFD) phase when cooled
below 105 K.\cite{Fleury1}  This transition, however, is observed to have little
effect on the dielectric properties. Cooling to still lower temperatures
results in a strong Curie-Weiss-like increase in the
static dielectric response, suggestive of a phase transition at
about 20 K. However, no transition actually occurs in that temperature
range; instead, the dielectric constant saturates to a
value of $\sim 2 \times 10^{4}$ at zero temperature.
\cite{Chaves1,Migoni1,Muller1,Muller2,Martonak1,Zhong1,Zhong2}
The absence of the ferroelectric (FE) transition can be explained by
quantum fluctuations of the atoms about their centrosymmetric cubic
positions (i.e., the formation of a ``quantum paraelectric'' state).
\cite{Muller1,Muller2,Martonak1,Zhong1,Cowley1,Worlock1}
The proximity to a ferroelectric transition is also evident from
experiments showing that modest uniaxial stress is capable of inducing
ferroelectricity.\cite{Uwe1}
Several first-principles \cite{Zhong1,Zhong2,KingSmith1,LaSota1,LaSota2}
and classical Monte-Carlo simulations on an effective Hamiltonian
\cite{Zhong3} have already provided valuable insight into the
structural properties and temperature dependence of FE phase
transitions in bulk STO. In particular, the interaction between FE and AFD
instabilities in the bulk phase has been thoroughly studied from first principles 
by Sai and Vanderbilt.\cite{Sai1} Moreover, a previous phenomenological
study of the effects of epitaxial strain on STO thin films by Pertsev {\it et al.},
\cite{Pertsev2} based on a Landau theory fit to
experimental data from bulk phases, produced a rich
temperature-strain phase diagram and provided support for the
idea that the oxygen-octahedron rotations have little influence on the dielectric
response.

In this work, we compute the effects of the in-plane epitaxial
lattice-matching constraint on the dielectric response and tunability of SrTiO$_{3}$
using first-principles density-functional theory within the local-density
approximation (LDA). We neglect the AFD instabilities in all
calculations. We also restrict
our analysis to zero temperature, but for simplicity neglect the quantum fluctuations.
Thus, our focus will be on the ferroelectric soft mode and its coupling to strain.
Despite these restrictions, the fact that we use a first-principles approach means that
we do not have to rely on empirical Landau parameters as in
Ref.~\onlinecite{Pertsev2}.  Thus, we can confidently make predictions
under conditions that vary drastically from those under which the data
determining the parameters were obtained, allowing us to consider the
effects of large epitaxial strains and
finite electric fields.

There is one major limitation of our theory, connected with the fact
that the LDA tends to underestimate lattice constants.  Because compression
tends to suppress ferroelectricity, this means that our
LDA system is ``less ferroelectric'' than true STO at zero temperature.
Indeed, we report below a soft-mode frequency of about 75 cm$^{-1}$
and a dielectric constant of about 390 at zero temperature for
unstrained STO within the LDA, whereas the experimental zero-temperature
system is exquisitely close to the ferroelectric transition with
$\omega_{\rm soft}\simeq10$ cm$^{-1}$ and $\epsilon \simeq 20,000$.\cite{Uwe1}
However, raising the temperature also has the effect of suppressing the
ferroelectric instability, so that the experimental system {\it at room
temperature} is characterized by $\omega_{\rm soft}\simeq 90$ cm$^{-1}$
and $\epsilon \simeq 290$.  Thus, fortuitously, the zero-temperature
LDA system corresponds reasonably well with the real physical system at a
temperature near, or a bit below, room temperature.  Comparisons between
these systems must obviously be approached with caution, but in fact
we find good semiquantitative agreement for several of the physical
properties of these two corresponding systems as will be presented
below.

The manuscript is organized as follows.
In Sec.~II, we provide the details of our
approach and approximations used, including our handling of finite
electric fields. In Sec.~III, we present and discuss the results of our
calculations for epitaxially strained STO, with emphasis on the strain
dependence of the tunability
by finite electric fields. Our conclusions appear in
Section IV.

\section{\label{method}METHOD}

\subsection{\label{fpc}First principles calculations}

First-principles density-functional calculations are performed
within the local density approximation (LDA) as implemented in the
PWSCF package.\cite{Baroni1} The exchange-correlation energy is
evaluated using the Ceperley-Alder form with Perdew-Zunger
parameterization.  Vanderbilt ultrasoft
pseudopotentials\cite{Vanderbilt1} are used treating 10 electrons
as valence for Sr (4$s^2$4$p^6$5$s^2$), 12 electrons for Ti
(3$s^2$3$p^6$4$s^2$3$d^2$), and 6 for O (2$s^2$2$p^4$). To achieve
well-converged results for small changes in the lattice constant,
the electronic wavefunctions are expanded in plane waves up to a
kinetic energy of 50 Ry. All calculations are performed with a
$6\!\times\!6\!\times\!6$ Monkhorst-Pack {\bf k}-point mesh.
\cite{Monkhorst1} To establish minimum-energy configurations we
converged the Hellmann-Feynman forces acting on the atoms to less 
than 0.1 mRy/a.u. 
Density functional perturbation theory (DFPT) is then used to calculate
the phonon frequencies of the structural optimized systems.

\subsection{\label{constraints}Structural constraints}

The role of epitaxial strain on the structural properties of STO
is isolated by systematically seeking the
ground-state structure of five-atom unit cells of {\it bulk} STO
constrained to several different in-plane lattice constants,
differing from the theoretical cubic lattice constant by fractions
ranging between $-$3\% and +3\% in steps of 0.5\%, and allowing all
atomic positions and the perpendicular lattice constant to relax fully
until the energy is minimized. 
To locate the phase transition points, and for the electric field
calculations, the in-plane lattice constant was varied in smaller
steps of 0.1\% in the paraelectric region. The presence of the
strain necessarily lowers the symmetry of the cubic STO system to
tetragonal at most; further spontaneous symmetry reduction occurs
for certain ranges of lattice constant.  In the remainder of this
subsection, we provide specific details of each of the structures considered
in this work and introduce the relevant notation.

\subsubsection{Epitaxial strains near zero}

The zero-strain paraelectric phase of SrTiO$_{3}$ has the ideal
cubic $Pm\overline{3}m$ perovskite structure, in which the octahedral oxygen
atoms lie at Wyckoff 3c positions $(\frac{1}{2},\frac{1}{2},0)$,
a single Ti atom lies at the body-centered site 1b
$(\frac{1}{2},\frac{1}{2},\frac{1}{2})$, and the lone Sr cation is at
1a $(0,0,0)$.  Our calculations result in a theoretical lattice
constant $a$ of 7.285 a.u., i.e., $\sim$$\,1.1\,\%$ less than the
experimental value of 7.365 a.u.; this underestimate is expected
when using the LDA and is consistent with several previous LDA-based
studies.

For small tensile or compressive strains, the structure of STO remains
centrosymmetric, but the symmetry is lowered to tetragonal (space
group {\it P4/mmm}). In this ``pseudocubic'' phase,
the cations sit at Wyckoff positions 1a $(0,0,0)$
for Sr and 1d $(\frac{1}{2},\frac{1}{2},\frac{1}{2})$ for Ti.
There are two different Wyckoff positions for oxygen: 1c
$(\frac{1}{2},\frac{1}{2},0)$, which will be referred to as 
O$_{\perp}$ (in the Ti-O chain along [001]), 
and 2e $(0,\frac{1}{2},\frac{1}{2}),(\frac{1}{2},0,\frac{1}{2})$ 
which will be referred to as O$_{\parallel x}$ and O$_{\parallel y}$
and lie in the same (001) plane as Ti.

\subsubsection{Compressive epitaxial strains}

For large enough compressive epitaxial strains, the symmetry of the
tetragonal phase is found to be lowered further to non-centrosymmetric {\it P4mm}.
This is a ferroelectric tetragonal structure with polarization along
[001].  The Wyckoff positions are 1a $(0,0,z=0)$ for Sr, 1b
$(\frac{1}{2},\frac{1}{2},\frac{1}{2}+\Delta{Ti_{z}})$ for Ti, 1b
$(\frac{1}{2},\frac{1}{2},\Delta{O_{\perp z}})$, and 2c
$(\frac{1}{2},0,\frac{1}{2}+\Delta{O_{\parallel z}})$ for oxygens.
The presence of a zone-boundary instability 
associated with rotation of the oxygen octahedra has been discussed for
this phase in earlier work,\cite{Sai2,LaSota1} and will not be
considered further here.

\subsubsection{Tensile epitaxial strains}

\begin{figure}[t]
\includegraphics[width=8cm]{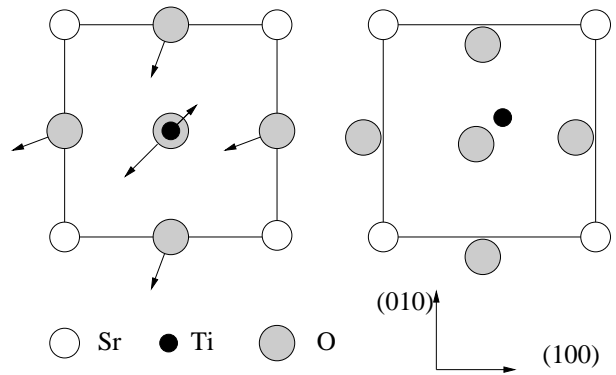}
\caption{\label{tensile-cell} 
The atomic displacements giving rise to the
ferroelectric {\it Amm2} structure in the tensile-strained
region are shown in a view of the primitive cubic unit cell projected
along [001].  
On the left: all
atoms at their ideal, centrosymmetric positions. Arrows indicate displacements 
of the Ti and oxygen atoms. The Sr atoms
are fixed at the corners of the cell.  Right: schematic view of the
{\it Amm2} structure (the displacements are exaggerated for clarity).}
\end{figure}

Above a critical value of tensile epitaxial strain, STO becomes
ferroelectric and transforms to an orthorhombic {\it Amm2}
structure.  This structure is non-centrosymmetric, exhibiting a nonzero
polarization along [110].  The Wyckoff positions associated
with this phase are 2a $(z=0,0,0)$ for Sr, 2b
$(\frac{1}{2}+\Delta{Ti_{x}},0,\frac{1}{2})$ for Ti, and 2b
$(\frac{1}{2}+\Delta{O_{\perp x}},0,0)$ and 4e
$(\frac{1}{4}+\Delta{O_{\parallel x}},\frac{1}{4}+\Delta{O_{\parallel y}},\frac{1}{2})$
for oxygens.
The structure is illustrated in Fig.~\ref{tensile-cell}.
The cell-doubling oxygen-octahedron rotation expected for this phase 
\cite{Pertsev2} will not be considered in the remainder of this paper.

\subsection{Electric field}

\subsubsection{\label{sec:fieldforce} Field-induced forces}

To describe the dielectric behavior of STO under finite dc bias, 
we must evaluate its properties in the presence of a homogeneous
electric field. This turns out to be subtle; fully 
first-principles methods for computing the behavior of periodic systems 
in finite fields 
have only recently been developed and are still in their nascent stages.
\cite{Souza1,Umari1} For STO, we resort to a 
simple and effective approximate technique that consists of first
computing the Born effective charge tensors $Z_i^{*}$ for each of
the ions $i$, and then adding a term $-e\,Z_i^{*}\,\bf\mathcal{E}$
to their Hellmann-Feynman forces ${\bf f}_i$,
where $e$ is the electronic charge and is defined to be positive.
The structure is relaxed until the total force on each atom is
close to zero, 
i.e., until
${\bf f}_i$ = $-e\,Z_i^{*}\,\bf\mathcal{E}$.  Similar approximations
were discussed by
Rabe\cite{Rabe1} and then implemented and used by Sai, Rabe and
Vanderbilt;\cite{Sai2} this approximation was also recently used by 
Fu and Bellaiche.\cite{Fu1}

To be more precise, we ideally would
determine the structure of STO by minimizing, with respect to
first wavefunctions $\psi_{n{\bf k}}$ and then to atomic coordinates ${\bf u}_i$, the
electric enthalpy $\mathcal{F}$ per unit cell,
\begin{equation}
\mathcal{F}({\bf\mathcal{E}}) = E_{KS} - \Omega
{\bf P} \cdot {\bf\mathcal{E}},
\end{equation}
where $E_{KS}$ is the internal energy obtained from a Kohn-Sham
functional (e.g., within the local-density or generalized-gradient
approximation),\cite{Souza1,explan-enthalpy}
${\bf P}={\bf P}_{\rm ion} +{\bf P}_{\rm el}$ is
the total (ionic plus electronic) macroscopic polarization,
${\bf\mathcal{E}}$ is the electric field, and $\Omega$ is the
volume of the unit cell.
$E_{KS}$ and $\bf P$ are explicit functions of atomic coordinates
${\bf u}_i$ and wavefunctions $\psi_{n{\bf k}}$, with the electric field
$\bf\mathcal{E}$ entering into $\mathcal{F}$ only through multiplication
of the functional $\bf P$. (Of course, $\mathcal{F}$, $E_{KS}$ and {\bf P}
can also be regarded as implicit functions of $\bf\mathcal{E}$ through
the dependence of the equilibrium values of the $\psi_{n{\bf k}}$ upon
$\bf\mathcal{E}$.)
By a Hellmann-Feynman argument,\cite{Souza1}
the total force acting on an ion is
\begin{equation}
-{\partial E_{KS}\over\partial{\bf u}_i} + 
\Omega {\partial{\bf P}\over\partial{\bf u}_i} \cdot {\bf\mathcal{E}}
= {\bf f}_i + {\rm Z}^*_i e \cdot {\bf\mathcal{E}},
\label{eq:two}
\end{equation}
where the ${\rm Z}^*_i$ are the Born effective charges; these also
depend implicitly on $\bf\mathcal{E}$, even at fixed ${\bf u}_i$,
via the $\psi_{n{\bf k}}(\bf\mathcal{E})$.

Here, we neglect the field dependence of ${\bf f}_i$
and ${\rm Z}^*_i$, computing these at ${\bf\mathcal{E}}=0$,  so
that the field enters only explicitly as the multiplier of
${\rm Z}^*_ie$ in Eq.~(\ref{eq:two}).   This is essentially the
approximation introduced in Ref.~\onlinecite{Sai2}.
Although we are able to compute the nonlinear behavior of the structural 
and dielectric properties insofar as they arise via field-induced
lattice displacements, we neglect purely electronic nonlinearities.
Thus, the results are rigorously correct only to first order in the field.
\cite{Sai2}

\subsubsection{\label{sec:subspace} Modeling in a reduced subspace}

In practice, we find that a straightforward minimization of $\mathcal F$
using the forces of Eq.~(\ref{eq:two}) leads to numerical instabilities
when the system is close to a second-order phase transition.  This
problem arises because the energy surface has a very shallow
minimum (or competing minima and saddle points) and the Hessian
matrix becomes poorly conditioned.  These numerical difficulties
can be solved by identifying a subspace that spans, to a good
approximation, the space of field-induced structural distortions,
and then parameterizing the energy in this subspace.
This modeling also helps us better understand the nonlinear
effects of the field on the structure and dielectric properties.

To identify the relevant subspace,
we begin by finding the pattern of atomic 
displacements produced by an infinitesimal electric field;
it is obtained
by multiplying the inverse of the force constant matrix
$\phi_{ij,\alpha\beta}$ with the forces $f_{i,\alpha}=-e\sum_\beta \
$Z$_{i,\alpha\beta}$$\mathcal{E}_\beta$. Explicitly, the displacement
$u_{i,\alpha}$ of atom $i$ in Cartesian direction $\alpha$ is
\begin{equation}
u_{i,\alpha}=-\sum_{j,\beta} \phi_{ij,\alpha\beta}^{-1} f_{j,\beta}.
\label{eq:udef}
\end{equation}
In this work we will consider electric fields applied exclusively along the 
$\hat{z}$ direction, 
perpendicular to the substrate assuming an (001) film, 
so the sum over $\beta$ above is
reduced to a single term.
We focus our attention in this part of the work on two
structures ($P4/mmm$ and $Amm2$); for these (as well as for $P4mm$, which
we will not consider further) the 
symmetry is such that the forces and displacements are also only along
$\hat{z}$.  Thus, Cartesian indices are dropped for the remainder of
this subsection, with all quantities referring implicitly to
$\hat{z}$ components only.

Once these displacements $u_i$ have been found, an expression for the electric enthalpy 
$\mathcal{F}({\bf \mathcal{E}})$ is obtained as follows.
We gather the $u_i$
into a $n$-dimensional vector ${\bf \xi}$ normalized to unity.
(Here $n$ is the number of atoms in the unit cell; $n=5$
for STO.) For each misfit strain, ${\bf \xi}$ defines
the subspace of possible field-induced displacements. 
The field dependence of the structure and polarization obtained in
this model are then that obtained by relaxing the 
ions and minimizing the electric enthalpy
subject to the constraint that the atomic displacements lie along
${\bf \xi}$. 
This should be a very good approximation for small displacements
considered here.

After constructing ${\bf \xi}$, we express the electric enthalpy in
terms of the scalar amplitude $u$. Keeping terms only to fourth order,
we obtain
\begin{eqnarray} \label{eq:enexp}
\mathcal{F}(u) &= &E_{0} + bu^{2} + du^{4} -\Omega\Delta P\mathcal{E} \nonumber\\
&= &E_{0} + bu^{2} + du^{4} - u \sum_{i}\xi_{i}\bar{Z}^{*}_{i}\mathcal{E}.
\end{eqnarray}
Here $\bar{Z}^{*}_{i}$ is the mode effective charge given below in
Eq.~(\ref{modeeffcharge}), computed at zero field and for the zero-field
structural parameters.
For each misfit strain, the amplitude $u$ was varied 
up to 0.03$\,c$ in steps of 0.002$\,c$ (where $c$ is the lattice
constant perpendicular to the implied substrate) to obtain the expansion 
coefficients $b$ and $d$ in Eq.~(\ref{eq:enexp}).\cite{detail1}

The electric field corresponding to a given $u$ is then extracted from
the equilibrium condition
\begin{equation}
\frac{\partial E}{\partial u} = 2bu + 4du^{3} - 
\sum_{i}\xi_{i}\bar{Z}^{*}_{i}\mathcal{E} = 0
\end{equation}
This leads to
\begin{equation} \label{eq:eu}
\mathcal{E}(u) = \frac{2bu+4du^{3}}{\sum_{i}\xi_{i}\bar{Z}^{*}_{i}}.
\end{equation}
Thus, having computed the quantities $P$, $\mathcal E$, and 
$\mathcal F$ on a mesh of $u$ values, we obtain parametric
relations between these quantities that can be plotted to
reveal the dielectric behavior of interest in a numerically
stable way.

\subsection{Dielectric response and tunability}

The dielectric function in the frequency range of the optical
phonons can be written as the sum of electronic and phonon 
contributions, that is,
\begin{equation}
\epsilon(\omega) = \epsilon^\infty + \epsilon^{\rm ph} (\omega).
\end{equation}
In most insulating perovskite oxides (ferroelectrics and related
materials), the electronic contribution is rather small 
($\epsilon^\infty \sim 5$)
and constant, and the static dielectric constant $\epsilon^0=\epsilon(0)$
is typically in the range of $20-100$, 
so the phonon
contribution is expected to dominate. In this work, we restrict our focus
to the static dielectric response and its tunability; we calculate and
analyze the phonon contribution, neglecting $\epsilon^\infty$.
For the remainder of the paper, we drop the superscripts 
and use the generic term dielectric response, even
though we only compute the phonon contribution. Thus, in
what follows, the dielectric constant $\epsilon$ has the meaning
of $\epsilon^{\rm ph}(0)$.

At zero bias, the zero-frequency phonon response is calculated in a 
straightforward manner using density functional perturbation theory 
(DFPT) to obtain the zone-center IR-active phonon modes and their 
frequencies, and using the Berry-phase theory of polarization\cite{King1} to 
compute Born effective charges by finite
differences. To evaluate the static response in an applied field,
and thus the tunability, we use the subspace approach
presented in Sec.~\ref{sec:subspace} above, which greatly simplifies
our treatment while retaining a high degree of accuracy.  In the
following, both methods are described in detail.

\subsubsection{\label{linresp} Zero dc bias}
In the absence of an electric field, the lattice contribution
to the static dielectric permittivity tensor $\epsilon_{0}$ can be written
\begin{equation}
\epsilon^{\rm ph}_{\alpha\beta} = \sum_m \frac{4\pi e^{2}}{M_{0}\Omega}
\frac{\tilde{Z}_{m\alpha}^{*}\tilde{Z}_{m\beta}^{*}}
{\omega_{m}^{2}} ,
\label{eq:phonon_epsilon}
\end{equation}
which includes a contribution 
from each of the zone-center polar modes $m$.
Here, $\Omega$ is the volume of the primitive unit cell, $M_{0}$ is
a reference mass taken as 1 amu, and $\tilde{Z}_{m\alpha}^{*}$ is the
mode effective charge,
\begin{equation}
\label{modeeffcharge}
\tilde{Z}_{m\alpha}^{*} = \sum_{i\gamma} Z_{\alpha\gamma}^{*}(i)
\,\sqrt{\frac{M_{0}}{M_{i}}} \; \hat{e}_{m}(i\gamma) ,
\end{equation}
where $\hat{e}_{m}(i\gamma)$ is a dynamical matrix eigenvector.
The corresponding real-space eigendisplacement of atom $i$ along
$\gamma$ is given by
$\hat{u}_{m}(i\gamma)=\hat{e}_{m}(i\gamma)/\sqrt{M_{i}}$.  The Born
effective charge $Z_{\alpha\gamma}^{*}(i)$ is given by
\begin{equation}
Z_{\alpha\gamma}^{*}(i) = \frac{\Omega}{|e|}
\;\frac{\partial P_{\alpha}}{\partial u_{i \gamma}} .
\end{equation}
In practice, we evaluate the effective charges by finite differences,
computing the change in polarization $\Delta P$ induced by several small
mode amplitudes $\Delta u_{i \gamma}$ via the Berry-phase approach using
a $6\times6\times20$ {\bf k}-point mesh.

\subsubsection{\label{efield}Nonzero dc bias}

To calculate the lattice contribution to the static dielectric constant 
in an applied electric field, we use the subspace approach presented in
Sec.~\ref{sec:subspace}. The field-induced change in the structure is
specified by a single parameter $u$, which determines the atomic displacements
through the normalized displacement vector $\xi$.
We express the lattice contribution to the dielectric susceptibility
in terms of the change in polarization induced by an applied electric field,
\begin{eqnarray} \label{eq:suscc}
\chi(\mathcal{E})&=&\frac{d P(\mathcal{E})}
{d\mathcal{E}} 
 =\frac{d P}{d u}\,\frac{d u}{d \mathcal{E}} 
 \nonumber\\
 &=& \Big(\sum_{i}\xi_{i}\bar{Z}^{*}_{i}\Big) \, \frac{d u}{d \mathcal{E}}
.
\end{eqnarray}
The expression $\epsilon=1 + 4\pi\chi(\mathcal{E})$ is used to
convert susceptibility to dielectric constant.
In practice, the derivative $d u / d \mathcal{E}$ is calculated 
numerically once $\mathcal{E}(u)$ is determined from Eq.~(\ref{eq:eu}). 
For strains close to the phase boundary, the zero-field dielectric constant computed in this 
manner is identical, by construction, to that obtained in Sec.II.D.1
using DFPT to obtain
phonon frequencies and eigenvectors, and the Berry-phase
calculations to obtain Born effective charges. 
Furthermore, for strains where the parameter $b$ was fit to energies for finite $u$,
as described in Sec. II.C.2, the zero-field dielectric constant is in excellent agreement with the
DFPT results.
To obtain the tunability, we compute the dielectric response for 
an appropriate range of values of $\mathcal{E}(u)$.

\section{\label{results}RESULTS AND DISCUSSION}

\subsection{\label{structural}Response to epitaxial strain}

\subsubsection{Structural properties}

As described in Sec.~\ref{constraints}, we first find the minimum-energy
structure of SrTiO$_3$ for values of the misfit strain between $-$3\%
and +3\%. For compressive strains less than $-$0.75\%, the lowest-energy
structure is ferroelectric 
tetragonal $P4mm$, with polarization along [001]. 
At $-$0.75\%, there is a continuous transition to the nonpolar
tetragonal $P4/mmm$ phase. At +0.54\%, there is another continuous
transition to the ferroelectric orthorhombic $Amm2$ structure, with 
polarization along [110].

\begin{figure}[t]
  \includegraphics[width=8cm]{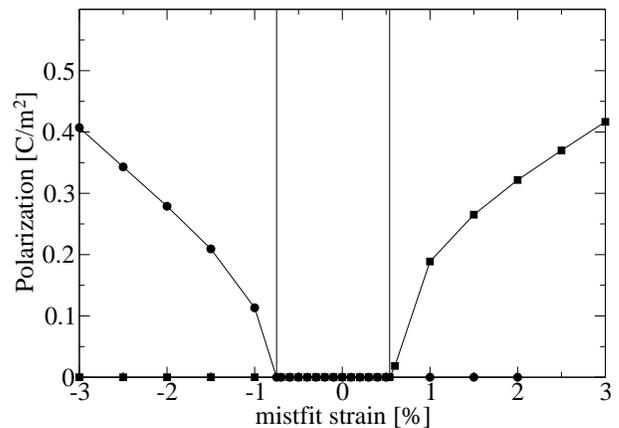}
\caption{\label{fig:polarization} Polarization as a function of
misfit strain. Solid circles and squares denotes polarization along [001]
and [110], respectively.}
\end{figure}

Figure \ref{fig:polarization} shows the polarization along [001] and
[110] directions.  The polarization increases
dramatically with strain, and for large strains above 2\% (tensile
or compressive), the magnitude of the polarization becomes
comparable to that of bulk BaTiO$_3$, a prototypical ferroelectric.
This suggests that by simply choosing the appropriate substrate,
the polarization of thin ferroelectric STO films, when under short-circuit
electrical boundary-conditions, could be tuned to a wide range of
values.

\begin{figure}[t]
  \includegraphics[width=8cm]{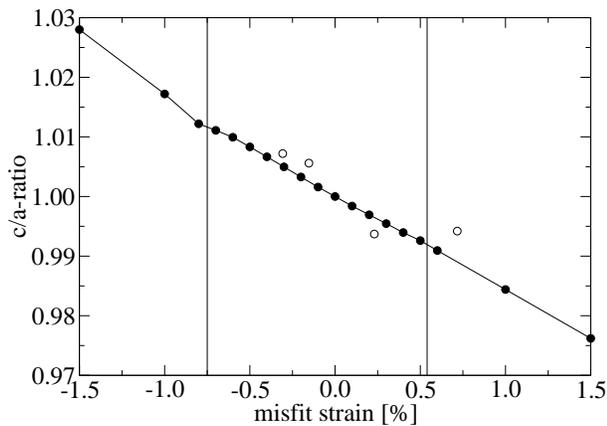}
\caption{\label{fig:ca} Calculated $c/a$ as a function of misfit
strain in SrTiO$_{3}$. The behaviors for $c/a < -$1.5\% and
$c/a > +$1.5\% (not shown) are very nearly given by linear extrapolation.
The open circles denote the samples measured by Hyun and
Char (Ref.~\protect\onlinecite{Hyun1}).}
\end{figure}

In  Fig.~\ref{fig:ca} we show the $c/a$ ratio as a function
of misfit strain.  For tensile (positive) misfit strains, $c/a$
drops almost linearly as the magnitude of the misfit strain
increases and is largely unaffected by the development of the in-plane
ferroelectric instability at the transition to the {\it Amm2}
phase.  For compressive epitaxial strains, in contrast, a
noticeable nonlinearity emerges when approaching the transition to
the {\it P4mm} phase; this may be indicative of the mounting structural
frustration that is eventually relieved by the occurrence of the
transition.  Once the phase boundary has been crossed, the onset and
growth of the $c$-axis polarization in the {\it P4mm} phase further increases
the $c/a$ ratio, and $c/a$ once again becomes very nearly
linear with misfit strain in the strongly compressive regime.
Also shown in Fig.~\ref{fig:ca} are some measurements of Hyun and
Char\cite{Hyun1} at 77 K that, while noisy, show a trend that is
roughly consistent with the theory, as will be discussed further in
Sec.~\ref{sec:dielprop}.

It is of interest to compare our misfit phase diagram with that
of Pertsev {\it et al.}\cite{Pertsev2,Pertsev3} The latter is
obtained by expressing the free energy as a
function of misfit strain, temperature, polarization, and
several additional order parameters corresponding to the linear
oxygen displacements that account for possible rotations of the
oxygen octahedra. The parameters in this Landau-theory expression are
obtained phenomenologically.

Regarding the nature of the polarization in the sequence of phases at
zero temperature, the present first-principles results are in 
very good agreement with the
Landau analysis. There are several detailed differences, however. 
First, our window of strains
over which the system remains paraelectric (between $-$0.75\% and +0.54\%)
is noticeably wider than that previously obtained (between $-$0.2 
and $-$0.02\% in Ref.~\onlinecite{Pertsev2,Pertsev3}).
Our overestimate of the stability of the paraelectric phase can
largely be attributed to the LDA. Specifically,
since our zero-temperature calculations within the LDA result in a slightly
smaller lattice constant than experiment, and since the smaller
volume stabilizes the paraelectric phase, our description of the
bulk system yields a more stable paraelectric phase, as our
underestimate of the bulk static dielectric constant attests
(calculated $\sim \! 400$ compared to the $\sim \! 2 \! \times \!
10^{4}$ observed experimentally).  In addition, 
octahedral rotations, not included in our analysis, might also alter the critical strain
corresponding to the FE phase transitions in this region.  


%
Another difference between the present work and
Refs.~\onlinecite{Pertsev2,Pertsev3}
is the prediction in the latter of a low-temperature
[100]-polarized ferroelectric phase under in a narrow window of tensile strain. 
Our results are consistent with only 
a single phase in this strain regime, with a polarization along [110].
Their polar [100]-oriented 
orthorhombic phase might be stabilized by the rotations not included here,
but since it occurs only in a very small strain region, we did not investigate
it further. For larger tensile strains, Pertsev {\it et al.}\cite{Pertsev3} predicted
a phase where the polarization is directed along [110] (like ours), with
an octahedral rotation around the same axis. 

\subsubsection{\label{dielectric} Dielectric properties}

In this section, we compute the misfit strain dependence of the 
lattice contribution to the dielectric response $\epsilon$ in zero electric field;
finite electric fields will be considered in the next section.
First, we discuss separately the zone-center phonons
and the Born effective charges that together determine $\epsilon$ according
to Eq.~(\ref{eq:phonon_epsilon}).

\begin{figure}[t]
  \includegraphics[width=8cm]{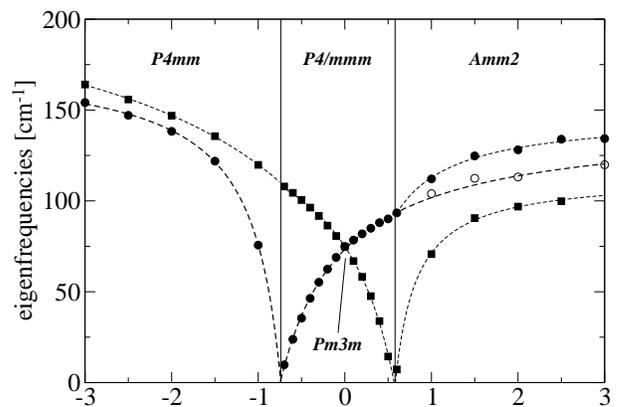}%
\caption{\label{fig:freq} FE soft-mode frequencies as a function of
misfit strain.  Solid circles indicate the FE soft mode polarized along [001].
Solid squares indicate the FE soft mode polarized along [110]. 
The vertical lines indicate the
phase transition points at $-$0.75\% and +0.54\% misfit strain.
Open circles denote the FE soft-mode frequency if the
polarization along [110] is suppressed by keeping the atoms on their centrosymmetric positions.}
\end{figure}

The zone-center phonons of each phase are calculated at each misfit strain using density-functional
perturbation theory. The lowest-frequency (softest)
polar modes dominate the dielectric response, and we show the frequencies 
of the softest in-plane and
out-of-plane transverse zone-center optical phonons as a function of misfit
strain in Fig.~\ref{fig:freq}. Their behavior reflects the phase
transition sequence discussed in the previous section.  
By symmetry, these modes are degenerate at
zero misfit strain.
However, the lowest-frequency (or soft) mode polarized
along [001] softens to zero as the misfit strain approaches
the critical value of $-$0.75\%, signaling the second-order transition
from the paraelectric {\it P4/mmm} to the [001]-polarized {\it P4mm}
structure.  Similarly, the lowest-frequency in-plane soft mode softens to zero at +0.54\%
misfit strain, marking the second-order transition from {\it P4/mmm} to {\it Amm2}, 
which has its FE polarization along the [110]
direction.  As can be seen in the figure, the in-plane soft
mode is only weakly affected at the {\it P4/mmm--P4mm} transition, while
the out-of-plane soft mode shows a significant hardening in the {\it Amm2} phase.
The hardening is produced by coupling to the in-plane polarization that develops in the 
{\it Amm2} phase. If the polarization is suppressed by keeping the atoms
on their centrosymmetric positions, the mode evolves smoothly with increase of the in-plane lattice
constant through the phase boundary, as is
shown by the open circles in  Fig.~\ref{fig:freq}.

\begin{figure}[t]
 \includegraphics[width=8cm]{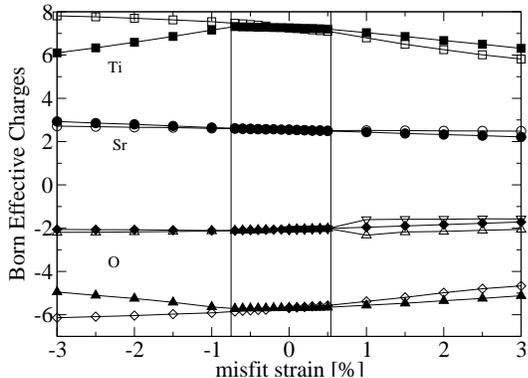}
\caption{\label{fig:zstars} Calculated Born effective charges
$Z^{*}_{33}$ (full symbols) and $Z^{*}_{11}$ (open symbols)
as a function of misfit strain.  The cubic-structure
values (at zero misfit strain) are $Z^{*}_{33}$(Ti)=7.25, $Z^{*}_{33}$(Sr)=2.55,
$Z^{*}_{33}$(O$_1$)=$-$2.06, and $Z^{*}_{33}$(O$_3$)=$-$5.69.
For O atoms,  up-triangles represent O$_3$, down-triangles O$_2$,
and diamonds O$_1$.}
\end{figure}

We now examine the Born effective charge tensors $Z^{*}_{i}$ as a
function of misfit strain.  In Fig.~\ref{fig:zstars} we present, for
brevity, only the $Z^{*}_{11}$ and $Z^{*}_{33}$ components of the
tensors, where `1' and `3' refer to the [100] and [001] directions, 
respectively.  While the Born effective
charge tensors are not diagonal in the Cartesian frame
in the {\it Amm2} phase (whose principal axes, by symmetry, are
along [110], [1$\bar 1$0], and [001]), the computed off-diagonal
components (not shown) are found to be quite small.
In what follows, O$_1$, O$_2$, and O$_3$ refer to the oxygen
atoms forming Ti-O chains in the $\hat{x}$, $\hat{y}$, and $\hat{z}$
directions, respectively.

Figure \ref{fig:zstars} shows that
in the paraelectric region, the Born effective charges are very close
to the values for the cubic structure, 7.25 for $Z^{*}_{33}$(Ti)
and $-$5.69 for $Z^{*}_{33}$(O$_3$).  As is well known, the fact that
these are anomalous (in the sense of exceeding the nominal valence)
arises from the hybridization of Ti and O orbitals in the Ti-O$_{3}$
chains and is quite sensitive to polar distortions of the
chain.\cite{Zhong4} Thus, in
the ferroelectric phases, we expect a significant misfit strain
dependence of the Born effective charges for
these atoms.  The effective charge of Ti, $Z^*_{33}$(Ti),
drops by almost 16\% in the ferroelectric region for compressive strain, with a
corresponding increase of $Z^*_{33}$(O$_3$). 
In comparison, $Z^{*}$(Sr) is rather insensitive to misfit strain over the whole
region, showing a weak trend towards lower values with
increasing misfit strain.  

In the tensile-strain-induced {\it Amm2} ferroelectric phase,
the orthorhombic symmetry (see Fig.~\ref{tensile-cell}) implies that
$Z^{*}_{11}$(O$_1$) = $Z^{*}_{22}$(O$_2$) $\ne$
$Z^{*}_{11}$(O$_2$) = $Z^{*}_{22}$(O$_1$).
In the range of strains shown, $Z^*_{11}$(O$_1$) shows a continuous increase of 18\%, while 
the $Z^{*}_{11}$ for O$_2$ and O$_3$ split slightly.
The charge neutrality sum rule for the $Z^{*}$ is maintained by
a corresponding decrease for $Z^{*}_{11}$(Ti).

\begin{figure}[t]
\includegraphics[width=8cm]{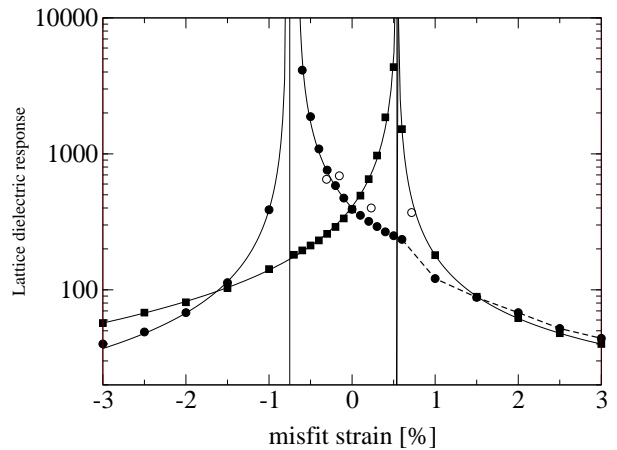}
\caption{\label{fig:epsilon} Dielectric constant as a function of
misfit strain.  Solid circles and squares denote $\epsilon_{33}$  and
$\epsilon_{11}$ respectively, where $3$ always denotes the 
[001] direction and $1$ refers to [100] in the {\it P4mm}
region and [110] in the {\it Amm2} region.  
Solid lines are
fits proportional to $(\eta-\eta_{\rm c})^{-1}$, where $\eta$ and
$\eta_{\rm c}$ are the actual and critical misfit strains, respectively.
Open circles are the $\epsilon_{33}$ measured by Hyun and Char
(Ref.~\protect\onlinecite{Hyun1}).}
\end{figure}

In Fig.~\ref{fig:epsilon}, we show the lattice contribution to the static 
dielectric response in zero electric field over the full range
of computed misfit strains.  The softening to zero frequency of the relevant
phonons at the second-order phase boundaries 
produces divergences
in the dielectric response near the critical strains. These
transitions show nearly perfect inverse-power-law behavior, except
at the points closest to the phase transitions where the low eigenfrequencies 
lead to numerical inaccuracies.  
Throughout the
paraelectric phase, both $\epsilon_{11}$ and $\epsilon_{33}$ are well over 100.
The abrupt drop of $\epsilon_{33}$ in the
tensile-strain region is related to the hardening of the
lowest-frequency phonon polarized along [001] shown in
Fig.~\ref{fig:freq}.
The comparison with the experimental data of Hyun and Char\cite{Hyun1}
will be discussed further in Sec.~\ref{sec:dielprop}.

\subsection{\label{tunability} Response to electric field}

Application of a finite electric field to epitaxially strained
STO leads to changes in the structure and dielectric response
that depend on misfit strain. In particular,
the sensitivity to applied field is expected to be largest near
the phase boundaries discussed above. In this section, we focus
our attention on the strain regime corresponding to
the zero-field paraelectric phase, with particular interest in
the behavior as the transition to the $P4mm$ phase is approached.
This regime is most relevant to practical applications, as the
ferroelectric state, its associated hysteresis, and the presence of
ferroelectric domains are generally undesirable for tunable device
applications.  We also consider electric fields
only along the [001] direction, in which case all displacements
also occur only along [001].

To briefly re-cap our approach as presented in Sec.~\ref{sec:subspace}, we
parametrize the electric field dependence of the atomic
displacements up to fourth order.  Combined with a linearized form
of the functional dependence of the polarization on atomic displacement,
this allows us to express the
lattice contribution to the dielectric susceptibility in terms
of the change in atomic displacements with electric field.
We first discuss the displacements induced by the applied
electric field, and then, in the following subsection, the resulting lattice contribution 
to the static dielectric response.

\subsubsection{Structural properties}

Using the approximate treatment of electric fields described in
Sec.~\ref{sec:fieldforce}, we calculate the misfit-strain dependence of 
the displacement response to a
small electric field. 
The resulting
displacements, defined relative to the Sr atom, are then
normalized to give the vector $\xi$ reported in Table
\ref{displacements} for selected misfit strains.  
The O displacements in an electric field are relatively insensitive to 
misfit strain, while the Ti displacement in an electric field grows 
with in-plane compressive strain, reflecting the
fact that in-plane compression, which is accompanied by a substantial
expansion of the $c$ lattice constant (see Fig.~\ref{fig:ca}), leads to
an opening of the oxygen octahedron in the $z$-direction and compression 
in the $xy$ plane. 
The negative sign of the displacements of O atoms in a positive electric field 
results from the negative signs of their Born effective charges, while Ti
is expected to exhibit a positive displacement since its Born effective
charge is larger than that of Sr.


\begin{table}[t]
\caption{\label{displacements}Normalized displacement vector $\xi$
for selected strained states in the paraelectric phase.}
\begin{center}
\begin{tabular}{rrrr}
\hline
\hline
\vspace{-0.1in}
\hspace{0.3in} & \hspace{0.87in} & \hspace{0.87in} & \hspace{0.87in}$\,$\\
strain & $u$(Ti) & $u$(O$_3$) & $u$(O$_1$) \\
\hline
$-$0.7\%  &0.0929 &$-$0.5081 &$-$0.6055  \\
$-$0.5\%  &0.0839 &$-$0.5111 &$-$0.6049  \\
$-$0.3\%  &0.0726 &$-$0.5129 &$-$0.6049  \\
$-$0.1\%  &0.0616 &$-$0.5143 &$-$0.6049  \\
 0.0\%    &0.0571 &$-$0.5156 &$-$0.6045  \\
 0.1\%    &0.0518 &$-$0.5158 &$-$0.6047  \\
 0.3\%    &0.0424 &$-$0.5176 &$-$0.6043  \\
 0.5\%    &0.0325 &$-$0.5192 &$-$0.6039  \\
\hline
\hline
\end{tabular}
\end{center}
\end{table}

For the larger compressive strains in Table \ref{displacements}, we
find that our field-induced displacement vectors (Eq.~(\ref{eq:udef}) 
are very similar
to the atomic displacement patterns of the [001]-polarized soft mode 
at zero electric field.  For example, the normalized atomic displacements of 
our computed normalized soft-mode eigenvector
at $-$0.7\% (near the critical misfit
strain of $-$0.75\%) are (Ti, O$_3$, O$_1$) = (0.0910, $-$0.5103, $-$0.6081),
almost identical to the displacement vector $\xi$ at the same misfit
strain. This shows that close to the phase boundary the
structural response to an electric field is almost entirely dominated by the
soft mode.

\subsubsection{\label{sec:dielprop} Dielectric properties and tunability}

\begin{figure}
\includegraphics[width=8cm]{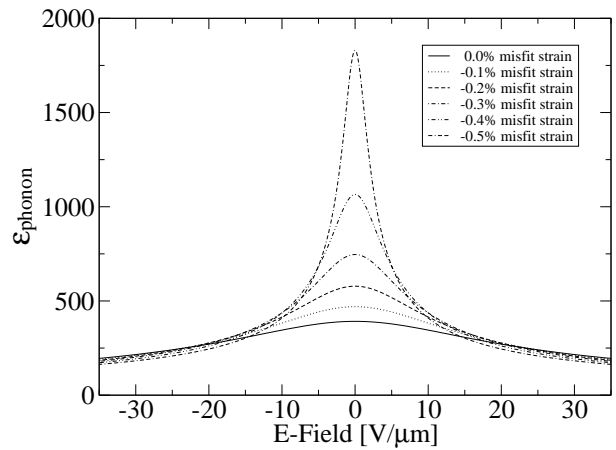}
\caption{\label{fig:evsfield} Dielectric constant vs electric
field for epitaxial STO in the paraelectric phase, for a series
of compressive strains approaching the transition to the
ferroelectric $P4mm$ phase.}
\end{figure}

We now investigate the
electric-field dependence of the lattice
contribution to the dielectric response along the [001] direction
as the in-plane compressive strain approaches the phase boundary
with the $P4mm$ phase. The results are shown
in Fig.~\ref{fig:evsfield}.  As the magnitude of the misfit strain
approaches the critical value of $-$0.75\%, the dielectric
response at low electric fields grows substantially.

We now gain further insight into our calculations through a quantitative
comparison with experiment. The curves in Fig.~\ref{fig:evsfield} resemble Lorentzians,
and this is expected from a phenomenological analysis, 
as follows.\cite{Devonshire1} Within the Landau-Devonshire
formalism, the ferroelectric phase transition can be
described by a free-energy functional $F$ expanded about the
paraelectric phase in even powers of the polarization $P$, i.e.,
\begin{equation}
F(P,T)=F_{0}+AP^{2}+BP^{4}+CP^{6}+\dots
\end{equation}
The coefficients depend on misfit strain and temperature; $A$ is generally
assumed to have the strongest dependence, while the variation of $B$ and 
higher order coefficients is small or even negligible.
Keeping only terms in $F$ to fourth order, the electric field is then given by
\begin{equation}
\mathcal{E}=\frac{\partial F}{\partial P}=2AP+4BP^{3},
\label{eq:efourth}
\end{equation}
and the dielectric susceptibility by
\begin{equation}
\label{eq:chifourth}
\frac{1}{\chi} = \frac{\partial \mathcal{E}}{\partial P} = 2A+12BP^{2}
\end{equation}
In the present discussion, the system is in the paraelectric 
phase ($A > 0$) so that these relations uniquely determine functions
$P({\mathcal E})$ and $\chi({\mathcal E})$.
From Eqs.~(\ref{eq:efourth}) and (\ref{eq:chifourth}), it is
easy to see that in this fourth-order approximation we expect
$\chi\rightarrow$ constant for $P\rightarrow0$, while
$\chi\propto\mathcal{E}^{-2/3}$ at large $P$.  A useful
approximate interpolation formula is then
\begin{equation}
\chi({\mathcal E})=\chi(0)
  \left[1+\left(\frac{\mathcal{E}}{\mathcal{E}_0}\right)^2\right]^{-1/3}
\end{equation}
as has been used previously in the literature.\cite{Auer1,Fuchs1}
Making the approximation $\epsilon\gg1$ so
that $\epsilon\simeq4\pi\chi$, we can write
\begin{equation}
\epsilon({\mathcal E})=\epsilon(0)
  \left[1+\left(\frac{\mathcal{E}}{\mathcal{E}_0}\right)^2\right]^{-1/3}
 \label{eq:epseq2}
\end{equation}
with $\epsilon(0)=4\pi\chi(0)$.

\begin{table}[t]
\caption{\label{table1}Zero-field dielectric constant $\epsilon(0)$ and
field scale $\mathcal{E}_0$ (reported as the slowly varying combination
$\epsilon(0)^{3/2}\mathcal{E}_0$) obtained from the fit of
Eq.~(\ref{eq:epseq2}) for each strain state in the paraelectric
compressively-strained region.}
\begin{center}
\begin{tabular}{rrr}
\hline
\hline
\vspace{-0.1in}
\hspace{0.5in} & \hspace{1.0in} & \hspace{1.7in}$\,$\\
strain & $\epsilon(0)\hspace{0.0in}$ & $\epsilon(0)^{3/2}\mathcal{E}_0 [V/cm]
   \hspace{0.0in}$ $\,$\\
\hline
 0.0\% &     391 &   8.3 \hspace{0.5in}$\,$\\
$-$0.1\% &     473 &   9.0 \hspace{0.5in}$\,$\\
$-$0.2\% &     584 &   9.3 \hspace{0.5in}$\,$\\
$-$0.3\% &     760 &   9.6 \hspace{0.5in}$\,$\\
$-$0.4\% &    1088 &  10.0 \hspace{0.5in}$\,$\\
$-$0.5\% &    1877 &  10.3 \hspace{0.5in}$\,$\\
$-$0.6\% &    4125 &  10.1 \hspace{0.5in}$\,$\\
$-$0.7\% &   19903 &  10.4 \hspace{0.5in}$\,$\\
\hline
\hline
\end{tabular}
\end{center}
\end{table}

A detailed analysis shows that $\mathcal{E}_{0}\propto
\epsilon_{0}^{-3/2}$, where the constant of proportionality is determined
by $B$, independent of $A$.  That is, $\epsilon(0)^{3/2}\mathcal{E}_0$
is not expected to depend strongly on proximity to the ferroelectric
phase transition, and thus is expected to vary only slowly with
temperature and misfit strain, and to be only weakly affected by the
LDA lattice-constant error.
Fuchs {\it et al.}\cite{Fuchs1} used Eq.~(\ref{eq:epseq2}) to fit
their experimentally-measured $\epsilon$-vs.-$\mathcal{E}$ data
at 200 K, and they observe dielectric constants between 1480 and 5270;
they find values of $\epsilon(0)^{3/2}\mathcal{E}_0$ ranging from
7.5 V/cm to 14.3 V/cm for films of 200-500 nm thickness. Fitting
our theoretical data to the same form, we extract values of
$\epsilon(0)^{3/2}\mathcal{E}_0$ ranging from 8.3 V/cm to 10.4 V/cm,
in general agreement with the findings of Fuchs {\it et al}. 
The results of our comparison are summarized in Table \ref{table1}.

Hyun and Char\cite{Hyun1} have grown SrTiO$_{3}$ on different
substrates to investigate the influence of epitaxial strain on the 
tunability of SrTiO$_{3}$.
For four samples, they report measurements at 77~K of in-plane
lattice constants, $c/a$ ratios, and dielectric constants,
as summarized in Table \ref{table3}.
\begin{table}[t]
\caption{\label{table3} Measured misfit strain, c/a, and zero electric field 
dielectric constant for STO films on various substrates.\protect\cite{Hyun1}
CRO and LAO stand for CaRuO$_3$ and LaAlO$_3$, respectively.}
\begin{center}
\begin{tabular}{rrrr}
\hline
\hline
\vspace{-0.1in}
\hspace{0.5in} & \hspace{0.6in} & \hspace{0.6in}$\,$ &  \hspace{1.3in}$\,$\\
strain & $c/a$ \hspace{0.0in} & $\epsilon(\mathcal{E} \! = \! 0)$ & materials\hspace{0.2in}$\,$ \\
\hline
$-$0.3073\% & 1.0072  & 640 & Au/STO/CRO/LAO\\
$-$0.1536\% & 1.0056  & 690 & Au/STO/CRO/STO\\
  +0.2305\% & 0.9937  & 400 & Au/STO/SRO/LAO\\
  +0.7170\% & 0.9942  & 360 & Au/STO/SRO/STO\\
\hline
\hline
\end{tabular}
\end{center}
\end{table}
There is good agreement between measured and calculated $c/a$ ratios for
the values of misfit strain observed in Ref.~\onlinecite{Hyun1},
as shown in Fig.~3.
For three of the samples, the dielectric constant increases with compressive misfit strain
in the paraelectric phase, in agreement with our calculations. 
The first sample is an exception: we expect it to exhibit a higher tunability, given its strain state.
This discrepancy can be attributed to the poorer quality of samples 
grown on LaAlO$_{3}$.\cite{Hyun1}  
Comparing our zero-temperature calculations in \ref{table3}) with
the measured value of 690 for the $-$0.1536\% sample, and an additional
computation of 319 for strain = +0.2\% with the measured value of 400 for the +0.2305\%
sample, (see Table \ref{table3}), we find surprisingly good agreement.
This may be partly due to slight shifts in the experimental
phase boundaries owing to finite temperature; 
because of our underestimate of lattice constants, our
paraelectric phase is over-stabilized, mimicking the effect
of temperature and shifting the calculated
phase boundary towards more compressive misfit strains.

\begin{figure}
 \includegraphics[width=8cm]{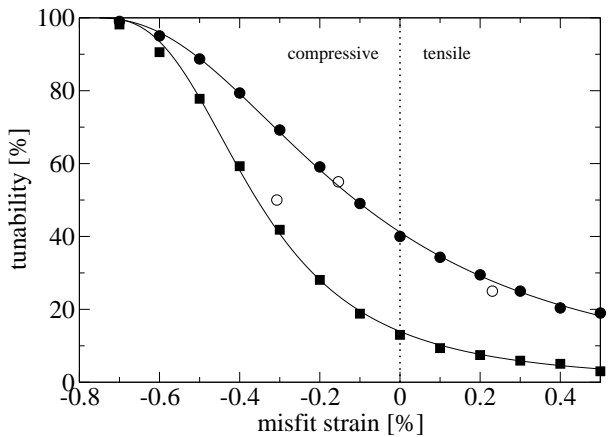}
\caption{\label{fig:tunability} Tunability 
$n=[\epsilon(0)-\epsilon(\mathcal{E})]/\epsilon(0)$ 
as a function of misfit strain. The values for $\epsilon(\mathcal{E})$ 
were taken at $\mathcal{E} = 10$ V/$\mu$m (full squares), and 
$\mathcal{E} = 25$ V/$\mu$m (full circles). Open circles are the
experimental results of Hyun and Char for $\mathcal{E} = 10$ V/$\mu$m.}
\end{figure}

Unlike the first three samples listed in Table \ref{table3}, which
we expect to be paraelectric based on their in-plane lattice constants,
the fourth sample should already be in the orthorhombic, tensile FE region,
according to our calculations and to the phase diagram of Pertsev
{\it et al.}  For this reason, we did not include this experimental
data point in Fig.~\ref{fig:tunability}.  We should like to note, however,
that the trends in the dependence of $c/a$ and $\epsilon(0)$ on in-plane
strain in Table \ref{table3} do not seem consistent regarding this last
data point.  As the strain changes from +0.23\% to
+0.72\%, the measured $c/a$ and $\epsilon(0)$ remain almost
identical.  The behavior of $c/a$, for example, seems inconsistent
with the roughly linear dependence of $c/a$ on misfit strain
predicted in our Fig.~\ref{fig:ca}, suggesting that more
thorough checks of the experimental behavior in the region of strong
tensile strain may be called for.

Finally, we have calculated the strain dependence of the tunability
parameter in the paraelectric phase for two different values of the bias
field, $\mathcal{E} = 10$ V/$\mu$m and $\mathcal{E} = 25$ V/$\mu$m, with
results shown in Fig.~\ref{fig:tunability}.
Although a direct comparison is not possible because the theoretical and
experimental conditions are rather different, the measurements of Hyun
and Char for $\mathcal{E} = 10$ V/$\mu$m are also shown.  These results
are closer to our calculated tunability for $\mathcal{E} = 25$ V/$\mu$m
than the one for $\mathcal{E} = 10$ V/$\mu$m, consistent with the fact
that our LDA-computed zero-temperature system is further from its
ferroelectric transition (as indicated, e.g., by the smaller zero-field
dielectric constant) than for the experimental system at 77 K, and
therefore less tunable.  Nevertheless, our results do tend to confirm
that a substantial variation in the tunability can be attained by growing
STO on substrates having different lattice constants.

\section{SUMMARY}

In this work, we examined the effects of in-plane biaxial strain
and an applied electric field on SrTiO$_3$ using first-principles
density-functional theory within the local density approximation.  
We computed the tunability of STO, and
found this quantity to be highly sensitive to epitaxial strain.  
We also find
that the dielectric constant itself varies significantly with
strain, and for sufficiently large compressive strains, STO can be made ferroelectric
with a polarization similar to that of bulk BaTiO$_3$. Our results
are in good agreement with available experiments.

Our studies complement and extend previous theoretical work on STO
films using a phenomenological Landau theory. Our parameter-free
description provides structural parameters and phonon frequencies
that can be compared with future experiments; in addition, we
observe significant differences between our zero-temperature phase
diagram and that obtained using Landau theories fit to empirical
data.

These results should be useful for the analysis of
ultrathin epitaxial films and superlattice structures
involving STO. In the latter, high strain states, achieved through layering,
can result in novel artificial materials with enhanced properties
over those of their bulk constituents, as was recently predicted for
BaTiO$_3$/SrTiO$_3$ superlattices.\cite{jeff}

For coherent epitaxial systems more complex than STO and at finite
temperatures, it seems clear that epitaxial strain will
generally be a crucial factor in determining properties.
In particular, we expect our results for STO to be a useful guide in understanding 
the strain dependence of more complex
ferroelectric thin films with high tunability, such as BST. 

\acknowledgments

This work was supported by ONR N00014-01-1-1061 and ONR N00014-00-1-0261 .

\begin{thebibliography}{00}

\bibitem{Fiedziuszko1}
S. J. Fiedziuszko, I. C. Hunter, T. Itho, Y. Kobayashi, T. Nishikawa,
S. N. Stitzer, and K. Wakino, IEEE Trans. Microwave Theory and Tech.
{\bf 50}, 706 (2002).

\bibitem{Varadan1}
V. K. Varadan, D. K. Gohdgaonkar, V.V. Varadan, J.F. Kelly, and P.
Glikerdas, Microwave J. {\bf Jan}, 116 (1992).

\bibitem{Horwitz1}
J. S. Horwitz, D. B. Chrisey, J. M. Pond, R. C. Y. Auyeung, C. M. Cotell,
K. S. Grabowski, P. C. Dorsey, and M. S. Kluskens, Integ.
Ferrolelectr. {\bf 8}, 53 (1995).

\bibitem{Yoneda1}
Y. Yoneda, K. Sakaue, and H. Terauchi, Surf. Sci. {\bf 529}, 283 (2003).

\bibitem{Qing1} 
W. Qing, W. Lian-wei, X. Shuo, S. Qin-wo, and L. Cheng-lu,
J. Funct. Mater. Devices. {\bf 8}, 128 (2002).

\bibitem{gim}
Y. Gim, T. Hudson, Y. Fan, C. Kwon, A. T. Findikoglu, B. J.
Gibbons, B. H. Park, and Q. X. Jia, Appl. Phys. Lett. {\bf 77},
1200 (2000).

\bibitem{knauss}
L. A. Knauss, J. M. Pond, J. S. Horowitz, D. B. Chrisey, C. H.
Mueller, and R. Treece, Appl. Phys. Lett. {\bf 69}, 25 (1996).

\bibitem{ramesh1}
C. L. Canedy, H. Li, S. P. Alpay, L. Salamanca-Riba, A. L.
Roytburd, and R. Ramesh, Appl. Phys. Lett. {\bf 77}, 1695 (2000).

\bibitem{Pertsev1}
N. A. Pertsev, A. G. Zembilgotov, and A. K. Tagantsev, Phys. Rev. Lett.
{\bf 80}, 1988 (1998).

\bibitem{James1}
A. R. James and X. X. Xi, J. Appl. Phys. {\bf 92}, 6149 (2002).

\bibitem{Park1}
B. H. Park, E. J. Peterson, Q.X. Jia, J. Lee, X. Zeng, W. Si, and
X. X. Xi, Appl. Phys. Lett. {\bf 78}, 533 (2001).

\bibitem{Ban1}
Z.-G. Ban and S.P. Alpay, J. Appl. Phys. {\bf 93}, 504 (2003).

\bibitem{Li1}
H. C. Li, W. Si, A.D. West, and X. X. Xi, 
Appl, Phys. Lett. {\bf 73}, 190 (1998).  

\bibitem{Hyun1}
S. Hyun and K. Char, Appl. Phys. Lett. {\bf 79}, 254 (2001).

\bibitem{Haeni1}
J.H. Haeni et al., unpublished.

\bibitem{Fleury1}
P. A. Fleury, J. F. Scott, and J. M. Worlock, Phys. Rev. Lett. {\bf
21}, 16 (1968).

\bibitem{Chaves1}
A. S. Chaves, F. C. S. Barreto, and L. A. A. Ribeiro, Phys. Rev. Lett.
{\bf 37}, 618 (1976).

\bibitem{Migoni1}
R. Migoni, H. Bilz, and D. B{\"a}uerle, Phys. Rev. Lett. {\bf 37},
1155 (1976).

\bibitem{Muller1}
K. A. M{\"u}ller and H. Burkard, Phys. Rev. B {\bf 19}, 3593 (1979).

\bibitem{Muller2}
K. A. M{\"u}ller, W. Berlinger, and E. Tosatti, Z. Phys. B: Condens.
Matter {\bf 84}, 277 (1991).

\bibitem{Martonak1}
R. Marto{\u{n}}{\'a}k and E. Tosatti, Phys. Rev. B {\bf 49}, 12596
(1994).

\bibitem{Zhong1}
W. Zhong and D. Vanderbilt, Phys. Rev. B {\bf 53}, 5047 (1996).

\bibitem{Zhong2}
W. Zhong and D. Vanderbilt, Phys. Rev. Lett. {\bf 74}, 2587 (1995).

\bibitem{Cowley1}
R. A. Cowley, Phys. Rev.  {\bf 134}, A981 (1964).

\bibitem{Worlock1}
J. M. Worlock, in {\it Structural Phase Transitions and Soft Modes},
edited by E.J. Samuelsen, E. Andersen, and J. Feder
(Universitetsforlaget, Oslo, 1971).

\bibitem{Uwe1}
H. Uwe and T. Sakudo, Phys. Rev. B {\bf 13}, 271 (1967).

\bibitem{KingSmith1}
R. D. King-Smith and D. Vanderbilt, Phys. Rev. B. {\bf 49}, 5828
(1994).

\bibitem{LaSota1}
C. LaSota, C. Z. Wang, R. Yu, and H. Krakauer, Ferroelectrics {\bf
194}, 109 (1997).

\bibitem{LaSota2}
C. LaSota, C.Z. Wang, R. Yu, and H. Krakauer, in {\it
First-Principles Calculations for Ferroelectrics: Fifth
Williamsburg Workshop}, edited by R.E. Cohen (AIP, Woodbury, NY,
1998), p. 139.

\bibitem{Zhong3}
W. Zhong, D. Vanderbilt, and K. M. Rabe, Phys. Rev. Lett. {\bf 73},
1861 (1994).

\bibitem{Sai1}
N. Sai and D. Vanderbilt, Phys. Rev. B {\bf 62}, 13942 (2000).

\bibitem{Pertsev2}
N. A. Pertsev, A. K. Tagantsev, and N. Setter, Phys. Rev. B {\bf 61},
R825 (2000).

\bibitem{Pertsev3}
N. A. Pertsev, A. K. Tagantsev, and N. Setter, Phys. Rev. B {\bf 65},
219901 (E) (2002).

\bibitem{Baroni1}
S. Baroni, A. Dal Corso, S. de Gironcoli, and P. Giannozzi,
\url{http://www.pwscf.org}.

\bibitem{Vanderbilt1}
D. Vanderbilt, Phys. Rev. B {\bf 41}, 7892 (1990).

\bibitem{Monkhorst1}
H. J. Monkhorst and J. D. Pack, Phys. Rev. B {\bf 13}, 5188 (1976).

\bibitem{Sai2}
N. Sai, K.M. Rabe, and D. Vanderbilt, Phys. Rev. B {\bf 66}, 104108
(2002).

\bibitem{Souza1}
I. Souza, J. Iniguez, D. Vanderbilt, Phys. Rev. Lett. {\bf 89},
117602 (2002).

\bibitem{Umari1}
P. Umari and A. Pasquarello, Phys. Rev. Lett. {\bf 89}, 157602 (2002).

\bibitem{Rabe1}
K. M. Rabe, MRS Proc. {\bf 718}, ed. by K. Poeppelmeier, A. Navrotsky and R. Wentzcovitch (2002).

\bibitem{Fu1}
H. Fu and L. Bellaiche, Phys. Rev. Lett. {\bf 91}, 57601 (2003).

\bibitem{explan-enthalpy}
To be precise, the electrostatic part of $E_{KS}({\bf\mathcal{E}})$
is defined here as the unit-cell integral of
$[\Delta\mathcal{E}_{\rm mic}({\bf r})]^2/8\pi$,
where $\Delta\mathcal{E}_{\rm mic}({\bf r})=
\mathcal{E}_{\rm mic}({\bf r})-{\bf\mathcal{E}}$ is the difference
between the microscopic and macroscopic fields.

\bibitem{detail1}
The accuracy of the lattice susceptibility when computed in this manner
depends on the fit of the energy expansion coefficients in Eq.~(\ref{eq:eu}).
At compressive strains near the ferroelectric phase transition, 
the energy as expressed in Eq.~(4) has a very small quadratic
coefficient $b$, and is thus increasingly difficult to fit reliably. 
This is especially problematic because, as can be seen from Eqs.~(\ref{eq:eu}) and (\ref{eq:suscc}), 
the dielectric constant in zero electric field 
depends only on the quadratic term in the energy expansion.
Thus, for compressive strains smaller than $-$0.4\%, we
adjust $b$ so that the zero-field dielectric constant obtained
from (11) agrees with that obtained from DFPT.
For these strain values, only the fourth order coefficient 
$d$ is fitted using the energy expansion.

\bibitem{King1}
R.D. King-Smith and D. Vanderbilt, Phys. Rev. B {\bf 47}, 1651 (1993). 

\bibitem{Zhong4}
W. Zhong, D. Vanderbilt, R. D. King-Smith and K. M. Rabe,
Ferroelectrics {\bf 164}, 291 (1995).

\bibitem{Devonshire1}
A. F. Devonshire, Philos. Mag. {\bf 40}, 1040 (1949).

\bibitem{Auer1}
R. Auer, E. Brecht, K. Herrmann, and R. Schneider, Physica C {\bf
299}, 177 (1998).

\bibitem{Fuchs1}
D. Fuchs, C. W. Schneider, R. Schneider, and H. Rietschel, J. Appl.
Phys. {\bf 85}, 7362 (1999).

\bibitem{jeff}
J. B. Neaton and K. M. Rabe, Appl. Phys. Lett. {\bf 82}, 1586 (2003).

\end{thebibliography}
\end{document}